\documentclass[aps,pra,preprint,10pt,a4paper]{revtex4-1}
\usepackage{amsfonts}
\usepackage{amssymb}
\usepackage{amsmath}
\usepackage{amsthm}
\usepackage{graphics}
\usepackage{graphicx}
\usepackage[colorlinks=true,linkcolor=blue,
urlcolor=blue,citecolor=blue]{hyperref}
\newtheorem{theorem}{Theorem}
\newtheorem{corollary}{Corollary}
\newtheorem{definition}{Definition}

\begin{document}
\title{Several nonlocal sets of multipartite pure orthogonal product states}

\author{Saronath Halder}
\email{saronath.halder@gmail.com}

\affiliation{Department of Mathematics, Indian Institute of Science Education 
and Research Berhampur, Transit Campus, Government ITI, Berhampur 760010, 
Odisha, India}

\begin{abstract}
It is known that there exist sets of pure orthogonal product states which 
cannot be perfectly distinguished by local operations and classical 
communication (LOCC). Such sets are nonlocal sets which exhibit nonlocality 
without entanglement. These nonlocal sets can be completable or uncompletable. 
In this work both completable and uncompletable small nonlocal sets of multipartite 
orthogonal product states are constructed. Apart from nonlocality, these 
sets have other interesting properties. In particular, the completable sets lead to 
the construction of a class of complete orthogonal product bases with the property 
that if such a basis is given then no state can be eliminated from that basis by 
performing orthogonality-preserving measurements. On the other hand, an 
uncompletable set of the present kind contains several Shifts unextendible product 
bases (UPBs) that belong to qubit subspaces. Identifying these subspace UPBs, it is 
possible to obtain a class of high-dimensional multipartite bound entangled states. 
Finally, it is shown that a two-qubit maximally entangled Bell state shared between 
any two parties is sufficient as a resource to distinguish the states of any completable 
set (of the above kind) perfectly by LOCC. This constitutes an example where the 
amount of entanglement, sufficient to accomplish the aforesaid task, depends neither 
on the dimension of the individual subsystems nor on the number of parties. 
\end{abstract}
\maketitle

\section{INTRODUCTION}\label{sec1}
In quantum information processing protocols, classical information is 
encoded within the state of a quantum system. Therefore, it is necessary 
to determine the state of a system in order to extract information. In 
this sense, after information encoding, it is required to discriminate 
among the possible states of a given system to decode that information. 
If the possible states of a given system are pairwise orthogonal to 
each other then the state of that system can in principle be determined 
perfectly by performing a suitable measurement on the {\it whole} system. 
Nevertheless, if the constituent parts (subsystems) of a composite 
quantum system are distributed among several spatially separated parties 
then it is not possible to perform measurements on the whole system. 
Such a constraint allows the parties to perform any sequence of quantum 
operations only on their individual subsystems and to make strategies 
they can communicate with each other classically. This class of 
operations on a distributed quantum system is commonly known as {\it 
local operations and classical communication} (LOCC). 

It is an established fact that the set of operations on a distributed 
quantum system which can be implemented by LOCC is a strict subset of 
all physically realizable quantum operations on the whole system. Thus, after 
information encoding the task of determining the state of a distributed 
quantum system perfectly by LOCC is not always possible even if the possible 
states of the system are pairwise orthogonal to each other. However, to study 
various properties of any composite quantum system, distributed among several 
spatially separated parties, LOCC plays a crucial role. Hence, to explore the 
properties of distributed quantum systems and to implement information 
processing tasks via distributed quantum systems, it is highly important to 
determine the states of such systems by LOCC. This particular task of 
determining the unknown state of a distributed quantum system by LOCC 
when a set of possible states is given is known as the LOCC state discrimination 
problem or local state discrimination problem (LSDP). 

The LSDP, as it is understood now, was first considered by Peres and Wootters 
\cite{Peres91}. During the last couple of decades, the LSDP has gotten considerable 
attention \cite{Walgate00, Virmani01, Ghosh01, Horodecki03, Fan04, Ghosh04, 
Nathanson05, Watrous05, Fan07, Duan07, Bandyopadhyay09, Yu12}. For a given 
set of pairwise orthogonal quantum states, if it is not possible to distinguish all 
the states perfectly by LOCC then the states are locally indistinguishable and 
the set is called a locally indistinguishable set or a {\it nonlocal} set. On the 
contrary, for a given set of pairwise orthogonal quantum states, if it is possible 
to distinguish all the states perfectly by LOCC then the states are locally 
distinguishable and the set is called a locally distinguishable set. If a nonlocal 
set forms a basis in a particular Hilbert space corresponding to a given quantum 
system then the set is said to be a nonlocal basis. Nonlocal sets have also found 
practical applications in data hiding \cite{Terhal01, Eggeling02}, quantum secret 
sharing \cite{Markham08}, etc. 

Due to the discovery of {\it quantum nonlocality without entanglement} by 
Bennett {\it et al.} \cite{Bennett99}, it is now understood that product states 
can also lead to local indistinguishability. In the paper just cited, the authors 
constructed two distinct nonlocal sets of orthogonal product states, forming 
complete bases in $\mathbb{C}^3\otimes\mathbb{C}^3$ and in $\mathbb{C}^2
\otimes\mathbb{C}^2\otimes\mathbb{C}^2$. Clearly, these bases constitute 
nonlocal separable operations, that is, separable operations which cannot be 
implemented by LOCC, though it is true that all quantum operations that can 
be implemented by LOCC are necessarily separable. In general, a nonlocal set 
of orthogonal product states that can be extended to a complete orthogonal 
product basis (COPB), always corresponds to a separable operation that cannot 
be realized by LOCC. In this context, it is important to mention that the mathematical 
structure of LOCC is still to be completely understood while separable operations are 
rich with mathematical structure. Again, if a quantum operation on a composite 
quantum system cannot be implemented by separable operations then that task must 
not be implemented by LOCC. This implies an important significance of exploring 
different types of separable operations. 

There are other types of sets of pure orthogonal product states, i.e., unextendible 
product bases (UPBs), uncompletable product bases (UCPBs), strongly uncompletable 
product bases (SUCPBs). The details regarding these sets can be found in Refs.~\cite{ Bennett99-1, DiVincenzo03}. These sets cannot be extended to COPBs. In fact, 
the orthogonal states of an unextendible product basis (UPB) or that of a strongly 
uncompletable product basis (SUCPB) cannot be perfectly distinguished by LOCC, 
whereas the orthogonal states within an uncompletable product basis (UCPB) 
cannot be perfectly distinguished by local projective measurements and classical 
communication \cite{DiVincenzo03}. Again, an important property of a UPB is that it 
leads to the generation of a bound entangled  state \cite{Bennett99-1, DiVincenzo03}, 
a mixed entangled state from which no pure  entangled state can be obtained by LOCC 
even if an arbitrary number of identical copies of the state are given. Examples of such 
states were first constructed in Ref.~\cite{Horodecki97}. Since then there has been no 
easy technique to detect bound entangled states. Therefore, constructing different 
classes of bound  entangled states is always a nontrivial task. In the following paragraph 
a few important results regarding different nonlocal sets of orthogonal product states 
are discussed. 

After the discovery of nonlocal COPBs by Bennett {\it et al.}, certain useful 
techniques were developed \cite{Groisman01, Walgate02} to prove the 
nonlocality of those COPBs. There are many other papers \cite{Rinaldis04, 
Niset06, Ye07, Feng09, Duan10, Yang13, Childs13, Zhang14, Yu15, Zhang15, 
Wang15, Chen15, Yang15, Zhang16, Xu16, Zhang16-1, Xu16-1, Wang17, 
Zhang17, Xu17, Wang17-1, Zhang17-1, Zhang17-2, Zhang17-3, Corke17, Halder18} 
in which nonlocal sets of orthogonal product states (OPSs) were constructed and 
different properties of such sets were studied. But multipartite systems are 
less explored with respect to the bipartite systems. This is in a sense that, apart 
from multipartite UPBs, mainly various structures of other nonlocal sets were 
studied. In Ref.~\cite{Niset06}, different classes of $m$-partite nonlocal sets of OPSs 
were constructed, where the dimension of any subsystem is dependent on the 
number of parties. Later \cite{Feng09, Yang13}, local distinguishability and 
indistinguishability of the OPSs which belong to $\mathbb{C}^2\otimes
\mathbb{C}^2\otimes\mathbb{C}^2$, were extensively studied. In Refs.~\cite{
Zhang14, Zhang15}, nonlocal COPBs were constructed, where the OPSs are 
associated with $\mathbb{C}^d\otimes\mathbb{C}^d\otimes\mathbb{C}^d$. 
Xu {\it et al.} showed that there exists a small nonlocal set of $m$-partite 
OPSs with only $2m$ members in $(\mathbb{C}^2)^{\otimes m}$ \cite{Xu16}. 
In Ref.~\cite{Wang17}, both bipartite and multipartite UPBs were presented. 
Furthermore, in Refs.~\cite{Zhang17, Xu17, Wang17-1, Zhang17-3}, other forms 
of nonlocal sets of multipartite OPSs were shown. However, in this work not only 
distinct nonlocal sets of multipartite OPSs are introduced but also a few interesting 
properties of such sets are investigated. 

Another direction of research is to explore entanglement as a resource to distinguish 
quantum states of nonlocal sets \cite{Cohen07, Cohen08, Bandyopadhyay09-1, 
Bandyopadhyay10, Duan14, Bandyopadhyay14, Bandyopadhyay16, Zhang16-2, 
Bandyopadhyay18, Zhang18} by LOCC. In particular, the problem of distinguishing 
product states of given nonlocal sets by LOCC using entanglement as a resource was 
considered in Refs.~\cite{Cohen08, Zhang16-2, Zhang18}. In Ref.~\cite{Cohen08}, 
Cohen constructed entanglement-assisted {\it local protocols} (protocols implementable 
by LOCC) to distinguish the orthogonal product states of several unextendible product 
bases. Later \cite{Zhang16-2}, separate entanglement-assisted local protocols were 
constructed to distinguish the states of a class of nonlocal bipartite COPBs. Recently 
\cite{Zhang18}, the problem of product state discrimination by LOCC was 
considered, where multiple copies of a maximally entangled state in $\mathbb{C}^2
\otimes\mathbb{C}^2$ are used as resource. Nonetheless, the present result 
of entanglement-assisted product state discrimination by LOCC is different form 
the existing results, and the resource state in this scenario is used quite efficiently. 

The remaining portion of this paper is arranged as follows. In Sec.~\ref{sec2}, 
necessary definitions and other preliminary concepts are presented. The forms of 
completable and uncompletable small nonlocal sets, associated with a three-qutrit, 
tripartite quantum system, are shown in Sec.~\ref{sec3}. In the same section, a 
few interesting properties of those sets are also discussed. Multipartite generalization 
of such sets is given in Sec.~\ref{sec4}. Next, in Sec.~\ref{sec5}, an 
entanglement-assisted product state discrimination protocol is constructed to 
distinguish the states of any completable set of the present kind by LOCC. Finally, 
the conclusion is drawn in Sec.~\ref{sec6} with some open problems for further 
studies.

\section{PRELIMINARIES}\label{sec2}
Before presenting the definitions, it is important to mention that, like other 
works on multipartite quantum systems \cite{Niset06, Feng09, Yang13, 
Zhang14, Zhang15, Xu16, Wang17, Zhang17, Xu17, Wang17-1, Zhang17-3}, 
in this work also only multipartite pure orthogonal fully separable states 
are considered and the states within a set are equally probable. The parties 
who are holding the subsystems are spatially separated and, thus, they are 
restricted to perform LOCC only. In a local discrimination protocol, there could 
be several rounds and, to complete such a protocol successfully, it is necessary 
to eliminate states of a given set by LOCC in several rounds. Here the product 
states are pairwise orthogonal to each other and thus it is quite relevant to 
consider the perfect discrimination of such product states. Note that in a 
multiround protocol it is necessary to preserve the orthogonality of the states 
after each round to distinguish the states perfectly.

In quantum theory, a measurement on a system of dimension $d$ can be 
expressed by a set of positive operator-valued measure (POVM) elements 
$\{\pi_l\}$. Such elements satisfy the completeness relation, that is, 
$\sum_l\pi_l$ = $\mathcal{I}_{d\times d}$, where $\mathcal{I}_{d\times d}$ 
is a $d\times d$ identity matrix. 

\begin{definition}\label{def1}
While distinguishing an unknown state of a given set, if all POVM elements of 
a measurement are proportional to the identity matrix then such a measurement 
is said to be a trivial measurement since such a measurement is not efficient 
to extract information useful for state discrimination. On the other hand, if not 
all POVM elements of a measurement are proportional to the identity matrix 
then the measurement is said to be a nontrivial measurement.
\end{definition}

\begin{definition}\label{def2}
Consider a measurement to distinguish a fixed set of pairwise orthogonal 
quantum states. After performing that measurement, if the postmeasurement 
states are also pairwise orthogonal to each other then such a measurement is 
said to be an orthogonality-preserving measurement. 
\end{definition}

Suppose a set of multipartite orthogonal quantum states is given. If no party is able 
to begin with a nontrivial orthogonality-preserving measurement then it is not possible 
to eliminate any state from the given set keeping the postmeasurement states orthogonal 
to each other. In fact, this guarantees nonlocality of the given set. This fact follows directly 
from the arguments given in Refs.~\cite{Groisman01, Walgate02}.

\begin{definition}\label{def3}
Consider an $m$-partite quantum system $\mathcal{H}$ = $\otimes_{i=1}^m
\mathcal{H}_i$. Also consider a set $S\in\mathcal{H}$ of pure orthogonal product 
states. The set $S$ constitutes a complete orthogonal product basis (COPB) if it 
spans $\mathcal{H}$ while the set $S$ is said to be an incomplete orthogonal 
product basis (ICOPB) if it spans a subspace $\mathcal{H}_S$ of $\mathcal{H}$. 
\end{definition}

\begin{definition}\label{def4}
An ICOPB in a fixed Hilbert space $\mathcal{H}$ constitutes an uncompletable product 
basis (UCPB) if the complementary subspace $\mathcal{H}_S^\perp$ contains a smaller 
number of pure orthogonal product states with respect to its dimension. On the other 
hand, an ICOPB is completable if the complementary subspace $\mathcal{H}_S^\perp$ 
is spanned by a set of orthogonal product states.
\end{definition}

\begin{definition}\label{def5}
Assume that an ICOPB is given. It is said to be an unextendible product basis (UPB) if 
there is no product state in the complementary subspace $\mathcal{H}_S^\perp$. 
\end{definition}

Consider a set $S$ of pure orthogonal product states $\{|\psi_k\rangle\}_{k=1}^n\in
\mathcal{H}$ = $\otimes_{i=1}^m\mathcal{H}_i$. Suppose, this set constitutes an 
unextendible product basis and it spans $\mathcal{H}_S$ of $\mathcal{H}$. Then, the 
normalized projector onto the complementary subspace $\mathcal{H}_S^\perp$ is 
given by-

\begin{equation}\label{eq1}
\rho = \frac{1}{D-n}(\mathcal{I}-\sum_{k=1}^n |\psi_k\rangle\langle\psi_k|),
\end{equation}

where $D$ is the net dimension of $\mathcal{H}$ and $\mathcal{I}$ is the identity 
operator onto $\mathcal{H}$. The state $\rho$ is an entangled state with positive 
partial transpose in each bipartition, that is, a multipartite bound entangled state 
\cite{Bennett99-1}. Next, the definition of another type of product basis is presented.

\begin{definition}\label{def6}
Suppose a UCPB is given in a fixed Hilbert space $\mathcal{H}$. This UCPB is said 
to be a strongly uncompletable product basis (SUCPB) if it cannot be completable in 
any locally extended Hilbert space $\mathcal{H}_{ext}$, where $\mathcal{H}_{ext}$ 
= $\otimes_{i=1}^m(\mathcal{H}_i\oplus\mathcal{H}_i^\prime)$. 
\end{definition}

Note that for a given Hilbert space a UPB is an obvious example of an SUCPB but an 
SUCPB may not be a UPB. However, here the definitions of UCPB, UPB, and SUCPB 
are given according to Refs.~\cite{Bennett99-1, DiVincenzo03}.

\section{TRIPARTITE SYSTEM}\label{sec3}
Consider a tripartite quantum system associated with a Hilbert 
space $\mathcal{H}$ = $(\mathbb{C}^3)^{\otimes3}$. In the following, 
an explicit construction of a small set $\mathcal{S}$ of pure 
orthogonal product states $\{|\psi_k\rangle\}\in\mathcal{H}$ = 
$(\mathbb{C}^3)^{\otimes3}$ is presented with $k=1,\dots,12$. To 
normalize the states of this section, consider $|a\pm b\rangle
\equiv(1/\sqrt{2})(|a\rangle\pm|b\rangle)$ for $a,b=0,1,2$. The 
states are given by-

\begin{equation}\label{eq2}
\begin{array}{ll}
|\psi_{1}\rangle =|0\rangle|1\rangle|0+1\rangle,  & 
|\psi_{2}\rangle =|0\rangle|1\rangle|0-1\rangle,  \\[0.5ex]

|\psi_{3}\rangle =|0\rangle|2\rangle|0+2\rangle,  &
|\psi_{4}\rangle =|0\rangle|2\rangle|0-2\rangle,  \\[0.5ex]

|\psi_{5}\rangle =|1\rangle|0+1\rangle|0\rangle,  & 
|\psi_{6}\rangle =|1\rangle|0-1\rangle|0\rangle,  \\[0.5ex]

|\psi_{7}\rangle =|2\rangle|0+2\rangle|0\rangle,  &
|\psi_{8}\rangle =|2\rangle|0-2\rangle|0\rangle,  \\[0.5ex]

|\psi_{9}\rangle =|0+1\rangle|0\rangle|1\rangle,  & 
|\psi_{10}\rangle=|0-1\rangle|0\rangle|1\rangle,  \\[0.5ex]

|\psi_{11}\rangle=|0+2\rangle|0\rangle|2\rangle,  &
|\psi_{12}\rangle=|0-2\rangle|0\rangle|2\rangle.
\end{array}
\end{equation}

Clearly, the set $\mathcal{S}$ does not form a COPB in $\mathcal{H}$ 
= $(\mathbb{C}^3)^{\otimes3}$ but it is possible to extend this set 
to a COPB by considering suitable product states (pairwise orthogonal). 
One such choice is given by-

\begin{equation}\label{eq3}
\begin{array}{lllll}
|0\rangle|0\rangle|0\rangle, & |0\rangle|1\rangle|2\rangle, & 
|0\rangle|2\rangle|1\rangle, & |1\rangle|0\rangle|2\rangle, & 
|1\rangle|1\rangle|1\rangle, \\[0.5ex]

|1\rangle|1\rangle|2\rangle, & |1\rangle|2\rangle|0\rangle, & 
|1\rangle|2\rangle|1\rangle, & |1\rangle|2\rangle|2\rangle, & 
|2\rangle|0\rangle|1\rangle, \\[0.5ex]

|2\rangle|1\rangle|0\rangle, & |2\rangle|1\rangle|1\rangle, & 
|2\rangle|1\rangle|2\rangle, & |2\rangle|2\rangle|1\rangle, & 
|2\rangle|2\rangle|2\rangle.
\end{array}
\end{equation}

The states of the set $\mathcal{S}$ and that of the above equation 
together form a COPB in $\mathcal{H}$ = $(\mathbb{C}^3)^{\otimes3}$. 
However, this set leads to an interesting property which is given 
in the following theorem. 

\begin{theorem}\label{th1}
Let $\mathcal{B}$ be a complete orthogonal product basis in a 
three-qutrit, tripartite Hilbert space. Then no state from $\mathcal{B}$ 
can be eliminated by performing orthogonality-preserving measurements 
if $\mathcal{B}$ contains $\mathcal{S}$.
\end{theorem}

\begin{proof}
To prove the above, first it is shown that the states of the set 
$\mathcal{S}$ allow each party to perform only trivial measurements 
on an entire three-dimensional subsystem if they want to preserve 
the orthogonality of the states.

Assume that the first party performs a measurement on his (or her) 
three-dimensional subsystem. Suppose this measurement is defined 
by a set of POVM elements $\{\pi_l\}$, $\sum_l\pi_l$ = 
$\mathcal{I}_{3\times3}$. The matrix form of $\pi_l$ = $M_l^\dagger M_l$ 
can be written in the \{$|0\rangle$, $|1\rangle$, $|2\rangle$\} basis 
and it is given by  

\begin{equation}\label{eq4}
\pi_l=M_l^\dagger M_l=
\begin{pmatrix}
e_{00} & e_{01} & e_{02} \\
e_{10} & e_{11} & e_{12} \\
e_{20} & e_{21} & e_{22} \\
\end{pmatrix}.
\end{equation}

If this measurement is orthogonality preserving then after 
the measurement by the first party the postmeasurement states 
remains pairwise orthogonal to each other. So, the states 
\{$(M_l\otimes\mathcal{I}\otimes\mathcal{I})|\psi_k\rangle,~ 
k = 1,\dots,12$\} must be orthogonal to each other. Setting 
the inner product of the postmeasurement states $(M_l\otimes
\mathcal{I}\otimes\mathcal{I})|\psi_5\rangle$ and $(M_l\otimes
\mathcal{I}\otimes\mathcal{I})|\psi_7\rangle$ equals to zero, it 
is found that $\langle1|M_l^\dagger M_l|2\rangle = e_{12} = 
\langle2|M_l^\dagger M_l|1\rangle = e_{21} =0$. Similarly, 
considering the inner product of the postmeasurement states 
$(M_l\otimes\mathcal{I}\otimes\mathcal{I})|\psi_1\rangle$ 
and $(M_l\otimes\mathcal{I}\otimes\mathcal{I})|\psi_5\rangle$, 
it turns out that $e_{01}$ = $e_{10}$ = 0. Again, the inner 
product of the postmeasurement states $(M_l\otimes\mathcal{I}
\otimes\mathcal{I})|\psi_3\rangle$, $(M_l\otimes\mathcal{I}\otimes
\mathcal{I})|\psi_7\rangle$ results in $e_{02}$ = $e_{20}$ = 0. 
In this way, it is proved that {\it all off-diagonal entries} 
of the above matrix are {\it zero}. 

Now, considering the inner product of the postmeasurement states 
$(M_l\otimes\mathcal{I}\otimes\mathcal{I})|\psi_9\rangle$ and 
$(M_l\otimes\mathcal{I}\otimes\mathcal{I})|\psi_{10}\rangle$ it 
is found that $\langle0+1|M_l^\dagger M_l|0-1\rangle$ = 0 and this 
implies that $e_{00}$ = $e_{11}$. Similarly, taking the inner 
product of $(M_l\otimes\mathcal{I}\otimes\mathcal{I})|\psi_{11}
\rangle$ and $(M_l\otimes\mathcal{I}\otimes\mathcal{I})|\psi_{12}
\rangle$, it turns out that $e_{00}$ = $e_{22}$. So, $e_{00}$ 
= $e_{11}$ = $e_{22}$, i.e., {\it all diagonal entries} of the 
above matrix are equal. 

So, the POVM elements $\{\pi_l\}$ that define the measurement 
for the first party are all proportional to a $3\times3$ identity 
matrix. This implies that the first party can perform only trivial 
measurements if he (or she) wants to preserve the orthogonality 
of the states. 

Notice that there is a symmetry present in the states of the set 
$\mathcal{S}$. For instance, if the order of the subsystems is 
rearranged in a way that the third party holds the subsystem of the 
first party, the second party holds the subsystem of the third party, 
and the first party holds the subsystem of the second party, then the 
states $|\psi_{1}\rangle$ and $|\psi_{2}\rangle$ are mapped to 
$|\psi_{5}\rangle$ and $|\psi_{6}\rangle$, respectively. Applying the 
same rearrangement rules, the following transformations are obtained: 
$|\psi_{5}\rangle\rightarrow|\psi_{9}\rangle$, 
$|\psi_{6}\rangle\rightarrow|\psi_{10}\rangle$, 
$|\psi_{9}\rangle\rightarrow|\psi_{1}\rangle$, 
$|\psi_{10}\rangle\rightarrow|\psi_{2}\rangle$, 
$|\psi_{3}\rangle\rightarrow|\psi_{7}\rangle$, 
$|\psi_{4}\rangle\rightarrow|\psi_{8}\rangle$, 
$|\psi_{7}\rangle\rightarrow|\psi_{11}\rangle$, 
$|\psi_{8}\rangle\rightarrow|\psi_{12}\rangle$, 
$|\psi_{11}\rangle\rightarrow|\psi_{3}\rangle$, and 
$|\psi_{12}\rangle\rightarrow|\psi_{4}\rangle$. 
Because of this symmetry, if one party cannot start with a nontrivial 
orthogonality-preserving measurement then the other parties cannot either. 
So, the states of the set $\mathcal{S}$ allow each party to start with 
only a trivial orthogonality-preserving measurement on their subsystems and 
this holds true for any basis $\mathcal{B}$ if $\mathcal{S}$ is fully 
contained in $\mathcal{B}$. 

Again, to eliminate any state from $\mathcal{B}$ by performing an 
orthogonality-preserving measurement, it is necessary that at least 
one party can start with a nontrivial measurement which also preserves 
the orthogonality of the states. In this sense, as no party can start 
with a nontrivial orthogonality-preserving measurement then it 
guarantees that no state from $\mathcal{B}$ can be eliminated by 
performing orthogonality-preserving measurements. This completes 
the proof.
\end{proof}

Obviously, the states of the set $\mathcal{S}$ constitute a sufficient 
condition for the basis $\mathcal{B}$ such that no state from this 
basis can be eliminated by performing orthogonality-preserving 
measurements. From the above discussion, it is also prominent that 
the states of $\mathcal{S}$ cannot be perfectly distinguished by 
LOCC. This is because of the fact that for local distinguishability it is 
necessary to eliminate state(s) of a given set, which is not possible 
in the present scenario. In this way, Theorem \ref{th1} provides a 
sufficient condition but not a necessary one for local indistinguishability 
of the states within the basis $\mathcal{B}$ in a three-qutrit tripartite 
Hilbert space. This particular notion is represented by the following 
corollary.

\begin{corollary}\label{coro1}
Let $\mathcal{B}^\prime$ be a complete orthogonal product basis in a 
three-qutrit tripartite Hilbert space. If no state from $\mathcal{B}^\prime$ 
can be eliminated by performing orthogonality-preserving measurements then 
the states of $\mathcal{B}^\prime$ cannot be perfectly distinguished by local 
operations and classical communication, though the converse is not always true. 
\end{corollary}

In support of the above corollary, a COPB $\mathcal{B}^\prime$ in $\mathcal{H} 
= (\mathbb{C}^3)^{\otimes3}$ is constructed from which certain states can be 
eliminated by performing an orthogonality-preserving measurement, though it is 
true that all the states of such a basis cannot be perfectly distinguished by LOCC.

Consider the states \{$|\psi_{1}\rangle$, $|\psi_{2}\rangle$, 
$|\psi_{5}\rangle$, $|\psi_{6}\rangle$, $|\psi_{9}\rangle$, 
$|\psi_{10}\rangle$\} of the set $\mathcal{S}$. It is known that 
these states cannot be perfectly distinguished by LOCC \cite{Xu16}. 
Thus, if a COPB $\mathcal{B}^\prime$ contains these states along 
with other product states then the basis must be a nonlocal basis. 
In order to construct such a basis $\mathcal{B}^\prime$ in 
$\mathcal{H}$ = $(\mathbb{C}^3)^{\otimes3}$, consider the following 
set of product states:

\begin{equation}\label{eq5}
\begin{array}{lll}
|0\rangle|0\rangle|0\rangle, & |0\rangle|0\rangle|2\rangle, & 
|0\rangle|1\rangle|2\rangle,\\[0.5ex]

|0\rangle|2\rangle|0\rangle, & |0\rangle|2\rangle|1\rangle, & 
|0\rangle|2\rangle|2\rangle,\\[0.5ex]

|1\rangle|0\rangle|2\rangle, & |1\rangle|1\rangle|1\rangle, & 
|1\rangle|1\rangle|2\rangle,\\[0.5ex]

|1\rangle|2\rangle|0\rangle, & |1\rangle|2\rangle|1\rangle, & 
|1\rangle|2\rangle|2\rangle,\\[0.5ex]

|2\rangle|0\rangle|0\rangle, & |2\rangle|0\rangle|1\rangle, & 
|2\rangle|0\rangle|2\rangle,\\[0.5ex] 

|2\rangle|1\rangle|0\rangle, & |2\rangle|1\rangle|1\rangle, & 
|2\rangle|1\rangle|2\rangle,\\[0.5ex] 

|2\rangle|2\rangle|0\rangle, & |2\rangle|2\rangle|1\rangle, & 
|2\rangle|2\rangle|2\rangle.
\end{array}
\end{equation}

The product states given above and the product states 
\{$|\psi_k\rangle$\} of $\mathcal{S}$ with $k = 1,2,5,6,9,10$ 
together form a COPB $\mathcal{B}^\prime$ in $\mathcal{H} = 
(\mathbb{C}^3)^{\otimes3}$. Interestingly, it is possible to 
define a nontrivial and orthogonality-preserving measurement by 
which the parties are able to eliminate certain states 
from $\mathcal{B}^\prime$. The states given in the previous 
equation excluding two states $|0\rangle|0\rangle|0\rangle$ and 
$|1\rangle|1\rangle|1\rangle$ can be eliminated from $\mathcal{B}
^\prime$. This can be done by performing a two-outcome projective 
measurement, and corresponding measurement operators are given 
by-

\begin{equation}\label{eq6}
\begin{array}{ll}
\pi_1 = |0\rangle\langle0|+|1\rangle\langle1|, & 
\pi_2 = |2\rangle\langle2|.
\end{array}
\end{equation}

This measurement can be performed by all three parties. In particular, 
if the measurement outcome is ``2'' due to the measurement by any of 
the parties then the state of the system can be perfectly determined by 
LOCC. Now, consider a three-qubit subspace $\mathcal{V}$ spanned by the 
following states:

\begin{equation}\label{eq7}
\begin{array}{ll}
|0\rangle|0\rangle|0\rangle,~|0\rangle|0\rangle|1\rangle,~ 
|0\rangle|1\rangle|0\rangle,~|0\rangle|1\rangle|1\rangle,\\[0.5ex]

|1\rangle|0\rangle|0\rangle,~|1\rangle|0\rangle|1\rangle,~ 
|1\rangle|1\rangle|0\rangle,~|1\rangle|1\rangle|1\rangle.
\end{array}
\end{equation}

Clearly, the states of the basis $\mathcal{B}^\prime$ that belongs to 
the subspace $\mathcal{V}$ cannot be perfectly distinguished by LOCC. 
This is solely because of the twisted states \{$|\psi_k\rangle$\} of 
$\mathcal{S}$ with $k = 1,2,5,6,9,10$. So, the basis $\mathcal{B}$ that 
contains $\mathcal{S}$ displays an additional feature besides just showing 
local indistinguishability of the states. 

Consider another small set $\mathcal{S}^\prime$. This set contains only 
seven pure orthogonal product states $\{|\psi_k\rangle\}\in\mathcal{H}$ 
= $(\mathbb{C}^3)^{\otimes 3}$ with $k$ = $1,\dots,7$. These states are 
given by-

\begin{equation}\label{eq8}
\begin{array}{c}
|\psi_1\rangle = |1\rangle|0-1\rangle|0+1\rangle,~
|\psi_2\rangle = |0+1\rangle|1\rangle|0-1\rangle,\\[0.5ex]

|\psi_3\rangle = |0-1\rangle|0+1\rangle|1\rangle,~
|\psi_4\rangle = |2\rangle|0-2\rangle|0+2\rangle,\\[0.5ex]

|\psi_5\rangle = |0+2\rangle|2\rangle|0-2\rangle,~
|\psi_6\rangle = |0-2\rangle|0+2\rangle|2\rangle,\\[0.5ex]

|\psi_7\rangle = |0\rangle|0\rangle|0\rangle.
\end{array}
\end{equation}

Notice that the states \{$|\psi_{1}\rangle$, $|\psi_{2}\rangle$, 
$|\psi_{3}\rangle$, $|\psi_{7}\rangle$\} ensure the fact that 
there is no other fully separable state which is orthogonal to 
these states and also belongs to the subspace $\mathcal{V}$ 
(defined earlier). In this sense, the states \{$|\psi_{1}\rangle$, 
$|\psi_{2}\rangle$, $|\psi_{3}\rangle$, $|\psi_{7}\rangle$\} 
together behave like a Shifts UPB \cite{Bennett99-1, DiVincenzo03} 
residing in the three-qubit subspace $\mathcal{V}$. If the states 
\{$|\psi_{k}\rangle$\} with $k$ = $1,2,3,7$ span $\mathcal{V}_S$ 
of $\mathcal{V}$ then $\mathcal{V}^\perp_S$ must be an entangled 
subspace, where $\mathcal{V}$ = $\mathcal{V}_S\oplus\mathcal{V}
^\perp_S$ and the normalized projector $\rho$ onto $\mathcal{V}
^\perp_S$ must be a tripartite bound entangled state. Now, 
consider another three-qubit subspace $\mathcal{V}^\prime$ which 
is spanned by-

\begin{equation}\label{eq9}
\begin{array}{l}
|0\rangle|0\rangle|0\rangle,~|0\rangle|0\rangle|2\rangle,~
|0\rangle|2\rangle|0\rangle,~|0\rangle|2\rangle|2\rangle,\\[0.5ex]

|2\rangle|0\rangle|0\rangle,~|2\rangle|0\rangle|2\rangle,~
|2\rangle|2\rangle|0\rangle,~|2\rangle|2\rangle|2\rangle.
\end{array}
\end{equation}

In the same way, as described before, the states \{$|\psi_{k}\rangle$\} 
with $k$ = $4,\dots,7$ together behave like a Shifts UPB residing 
in the three-qubit subspace $\mathcal{V}^\prime$. If the states 
\{$|\psi_{k}\rangle$\} with $k$ = $4,\dots,7$ spans $\mathcal{V}_S^
\prime$ of $\mathcal{V}^\prime$ then $\mathcal{V}^{\prime\perp}_S$ 
must be another entangled subspace, where $\mathcal{V}^\prime$ = 
$\mathcal{V}_S^\prime\oplus\mathcal{V}^{\prime\perp}_S$ and the 
normalized projector $\rho^\prime$ onto $\mathcal{V}^{\prime\perp}_S$ 
is another tripartite bound entangled state. Note that the subspaces 
$\mathcal{V}$ and $\mathcal{V}^\prime$ are not completely disjoint. 
In particular, $\mathcal{V}_S$ and $\mathcal{V}^\prime_S$ have 
one-dimensional overlap with each other. However, with the help of 
$\rho$ and $\rho^\prime$, it is possible to define a new class of 
tripartite bound entangled states given by-

\begin{equation}\label{eq10}
\sigma(\lambda)=\lambda\rho+(1-\lambda)\rho^\prime,
\end{equation}

where $\lambda$ takes the values between 0 and 1. The states 
$\sigma(\lambda)$ are supported in $\mathcal{V}_S^\perp\oplus
\mathcal{V}_S^{\prime\perp}$ which is an entangled subspace. 
Therefore, the states $\sigma(\lambda)$ must be entangled states. 
Again, a convex combination of two bound entangled states cannot 
lead to distillable entanglement. Hence, the states $\sigma(\lambda)$ 
must be bound entangled states. In this context it is important to 
mention that both $\rho$ and $\rho^\prime$ are separable in the 
bipartitions \cite{DiVincenzo03}. This results that the states 
$\sigma(\lambda)$ are also separable in the bipartitions. 

Clearly, it is not possible to extend the set $\mathcal{S}^\prime$ 
to a COPB in $\mathcal{H}$ = $(\mathbb{C}^3)^{\otimes 3}$ as this 
set $\mathcal{S}^\prime$ contains several UPBs residing in the subspaces   
$\mathcal{V}$ and $\mathcal{V}^\prime$. Thus, it is obvious that the 
states of $\mathcal{S}^\prime$ cannot be perfectly distinguished by 
LOCC. This is because of the fact that the states within a UPB cannot 
be perfectly distinguished by LOCC \cite{Bennett99-1}. Moreover, the 
set $\mathcal{S}^\prime$ constitutes an SUCPB because those subspace 
UPBs lead to the uncompletability in any locally extended Hilbert space.

In the next section, both the sets $\mathcal{S}$ and $\mathcal{S}
^\prime$ are generalized for an arbitrary number of parties with 
high-dimensional subsystems and then different properties of these 
small sets are explored.

\section{MULTIPARTITE SYSTEM}\label{sec4}
Consider a multipartite quantum system associated with a Hilbert 
space $\mathcal{H}$ = $(\mathbb{C}^d)^{\otimes m}$ ($d\geq3$, $m\geq3$), 
where $m$ is the number of parties and each party holds a $d$-dimensional 
subsystem. In the following, a set $\mathbb{S}$ of $2m(d-1)$ pure 
orthogonal $m$-partite product states in $\mathcal{H}$ = 
$(\mathbb{C}^d)^{\otimes m}$ is constructed. To normalize the states 
consider $|0\pm i\rangle\equiv(1/\sqrt{2})(|0\rangle + |i\rangle)$ 
for $i$ = $1,\dots,(d-1)$. The states are given by-

\begin{equation}\label{eq11}
\begin{array}{c}
|\psi_{1i}\rangle=|0\rangle|0\rangle\cdots|0\rangle|i\rangle|0+i\rangle, \\
|\psi_{1i}^\perp\rangle=|0\rangle|0\rangle\cdots|0\rangle|i\rangle|0-i\rangle, \\[1ex]

|\psi_{2i}\rangle=|0\rangle\cdots|0\rangle|i\rangle|0+i\rangle|0\rangle, \\
|\psi_{2i}^\perp\rangle=|0\rangle\cdots|0\rangle|i\rangle|0-i\rangle|0\rangle, \\[1ex]

\vdots \\[1ex]

|\psi_{mi}\rangle=|0+i\rangle|0\rangle|0\rangle\cdots|0\rangle|i\rangle, \\
|\psi_{mi}^\perp\rangle=|0-i\rangle|0\rangle|0\rangle\cdots|0\rangle|i\rangle. 
\end{array}
\end{equation}

It is possible to extend the set $\mathbb{S}$ to a COPB in $\mathcal{H}$ 
= $(\mathbb{C}^d)^{\otimes m}$. For this purpose, consider a different COPB 
$\mathbf{B}$ in $\mathcal{H}$ = $(\mathbb{C}^d)^{\otimes m}$. The form of 
the product states contained in $\mathbf{B}$ is given by 
$|b_1\rangle|b_2\rangle\cdots|b_m\rangle$, where $b_i$ = $0,\dots,(d-1)$. 
Next, consider a different set of orthogonal product states $\mathbf{S}$ in 
$\mathcal{H}$ = $(\mathbb{C}^d)^{\otimes m}$, given by-

\begin{equation}\label{eq12}
\begin{array}{c}
|0\rangle|0\rangle\cdots|0\rangle|i\rangle|0\rangle, ~
|0\rangle|0\rangle\cdots|0\rangle|i\rangle|i\rangle, \\ [1ex]

|0\rangle\cdots|0\rangle|i\rangle|0\rangle|0\rangle, ~ 
|0\rangle\cdots|0\rangle|i\rangle|i\rangle|0\rangle, \\ [1ex]

\vdots \\ [1ex]

|0\rangle|0\rangle|0\rangle\cdots|0\rangle|i\rangle, ~
|i\rangle|0\rangle|0\rangle\cdots|0\rangle|i\rangle,
\end{array}
\end{equation}

where $i$ = $1,\dots,(d-1)$. Notice that both the sets $\mathbb{S}$ and 
$\mathbf{S}$ span the same subspace $\mathcal{H}^\prime$ of $\mathcal{H}$ 
= $(\mathbb{C}^d)^{\otimes m}$. Now, define another set of product states 
as $\mathbf{S}^\prime$ = $\mathbf{B}-\mathbf{S}$. The states of the set 
$\mathbb{S}$ and that of $\mathbf{S}^\prime$ together form a COPB in 
$\mathcal{H}$ = $(\mathbb{C}^d)^{\otimes m}$. Make a note that if one puts 
$m$ = $d$ = 3 then the set $\mathbb{S}$ is exactly the same as that of 
$\mathcal{S}$ (defined in the previous section), only the states are labeled 
differently. However, the set $\mathbb{S}$ leads to an interesting property 
which is captured by the next theorem. This theorem can be regarded as the 
generalized version of Theorem \ref{th1}.

\begin{theorem}\label{th2}
Let $\mathbb{B}$ be a complete orthogonal product basis in an $m$-qudit, 
$m$-partite Hilbert space. Then no state from $\mathbb{B}$ can be eliminated 
by performing orthogonality-preserving measurements if $\mathbb{B}$ 
contains $\mathbb{S}$.
\end{theorem}

\begin{proof}
To prove the above, it is sufficient to show that the states of the set 
$\mathbb{S}$ allow each party to perform only trivial measurements on an 
entire $d$-dimensional subsystem if they want to preserve the orthogonality 
of the states. This argument follows from the proof of Theorem \ref{th1}. 

Now, assume that the $(m-1)$th party performs a measurement defined by a 
set of POVM elements $\{\pi_l\}$; $\pi_l$ = $M_l^\dagger M_l$ and $\sum_l\pi_l$ 
= $\mathcal{I}_{d\times d}$. Each $\pi_l$ can be represented by a $d\times d$ 
matrix written in the $\{|0\rangle,|1\rangle,\dots,|d-1\rangle\}$ basis as the 
following:

\begin{equation}\label{eq13}
\pi_l=M_l^\dagger M_l=
\begin{pmatrix}
e_{00} & e_{01} & \hdots & e_{0,d-1} \\
e_{10} & e_{11} & \hdots & e_{1,d-1} \\
\vdots & \vdots & \ddots & \vdots    \\
e_{d-1,0} & e_{d-1,1} & \hdots & e_{d-1,d-1} \\
\end{pmatrix}.
\end{equation}

If this measurement preserves the orthogonality of the states 
then the postmeasurement states must be pairwise orthogonal 
to each other. Now, setting the inner product $\langle\psi_{1i}
|\mathcal{I}\otimes\mathcal{I}\otimes\cdots\otimes M_l^\dagger M_l
\otimes\mathcal{I}|\psi_{1i^\prime}\rangle$ = 0, it is found 
that $\langle i|M_l^\dagger M_l|i^\prime\rangle$ = 0, or 
$e_{ii^\prime}$ = 0 with $i,i^\prime$ = $1,\dots,(d-1)$ and 
$i\neq i^\prime$. These $e_{ii^\prime}$ are the off-diagonal 
entries of the above matrix. Next, consider the states $|\psi_{1i}
\rangle$ and $|\psi_{mi}\rangle$. The inner product 
$\langle\psi_{mi}|\mathcal{I}\otimes\mathcal{I}\otimes
\cdots\otimes M_l^\dagger M_l\otimes\mathcal{I}|\psi_{1i}\rangle$ 
must be zero and it turns out that $\langle0|M_l^\dagger M_l|i\rangle$ 
= $e_{0i}$ = $e_{i0}$ = 0 for $i$ = $1,\dots,(d-1)$. Thus, all the 
off-diagonal entries of the above matrix are zero. 

Next, it is shown that the diagonal entries of the above matrix 
are all the same. For this purpose, consider the states $|\psi_{2i}
\rangle$ and $|\psi_{2i}^\perp\rangle$. Again, the inner product 
$\langle\psi_{2i}|\mathcal{I}\otimes\mathcal{I}\otimes\cdots\otimes 
M_l^\dagger M_l\otimes\mathcal{I}|\psi_{2i}^\perp\rangle$ = 0 and 
it is found that $\langle0+i|M_l^\dagger M_l|0-i\rangle$ = 0, or 
$e_{00}$ = $e_{ii}$. Hence, the diagonal entries of the above 
matrix are all the same. 

In this way, it is proved that the POVM elements that define the 
measurement for the $(m-1)$th party are proportional to a $d\times d$ 
identity matrix. So, the $(m-1)$th party performs only trivial 
measurements. 

Similarly, it can be shown that all other parties can perform only 
trivial measurements if they want to preserve the orthogonality of 
the states. This is because of the symmetry present in the states 
of the set $\mathbb{S}$. Here the proof completes. 
\end{proof}

For any COPB $\mathbb{B}\in\mathcal{H}$ = $(\mathbb{C}^d)^{\otimes m}$ 
that contains all the states of the set $\mathbb{S}$, it is straightforward 
from Theorem \ref{th2} that the states of $\mathbb{B}$ cannot be perfectly 
distinguished by LOCC. If the construction is restricted up to a bipartite 
system, then there exists a set of $4(d-1)$ product states that cannot 
be distinguished by LOCC. Precise construction of such a set is given in 
Ref.~\cite{Zhang15}. Again, Walgate {\it et al.} showed that there does not exist a 
product basis in $\mathbb{C}^2\otimes\mathbb{C}^2$ which is a nonlocal basis 
\cite{Walgate02}. When the number of parties $m\geq3$, and every party holds a 
qubit, then there are $2m$ product states in $(\mathbb{C}^2)^{\otimes m}$, that 
cannot be perfectly distinguished by LOCC. Such a class of nonlocal sets is 
constructed in Ref.~\cite{Xu16}. Next, the generalized version of the uncompletable 
set is presented.

Consider a set $\mathbb{S}^\prime$ of $m$-partite orthogonal product states in 
$\mathcal{H}$ = $(\mathbb{C}^d)^{\otimes m}$ ($d\geq3$, $m\geq3$). Note that 
here $m$ is an odd number and can be considered as $2p+1$. The states are given 
by-

\begin{equation}\label{eq14}
\begin{array}{c}
|\phi_{1i}\rangle = |i\rangle|\alpha_1\rangle|\alpha_2\rangle\cdots|\alpha_p\rangle
|\alpha_p^\perp\rangle\cdots|\alpha_2^\perp\rangle|\alpha_1^\perp\rangle,\\[1ex]

|\phi_{2i}\rangle = |\alpha_1^\perp\rangle|i\rangle|\alpha_1\rangle|\alpha_2\rangle
\cdots|\alpha_p\rangle|\alpha_p^\perp\rangle\cdots|\alpha_2^\perp\rangle,\\[1ex]

\vdots\\[1ex]

|\phi_{mi}\rangle = |\alpha_1\rangle|\alpha_2\rangle\cdots|\alpha_p\rangle|\alpha_p
^\perp\rangle\cdots|\alpha_2^\perp\rangle|\alpha_1^\perp\rangle|i\rangle,\\[1ex]

|\phi^\prime\rangle = |0\rangle|0\rangle\cdots|0\rangle.
\end{array}
\end{equation}

$i$ = $1,\dots,(d-1)$, and the length of the set is $m(d-1)+1$. The states $|\alpha_r\rangle$ 
and $|\alpha_s\rangle$ for all $r\neq s$ are chosen to be neither orthogonal nor identical. 
Also, $|\alpha_r\rangle$ is a linear combination of the states $|0\rangle$ and $|i\rangle$ 
for all $r$. Clearly, for a fixed value of $i$, the states \{$|\phi_{1i}\rangle, |\phi_{2i}\rangle,
\cdots, |\phi_{mi}\rangle, |\phi^\prime\rangle$\} forms an $m$-partite Shifts UPB residing 
in a $2^m$-dimensional subspace spanned by -

\begin{equation}\label{eq15}
\begin{array}{c}
|0\rangle\cdots|0\rangle|0\rangle,~~ |0\rangle\cdots|0\rangle|i\rangle,\\
|0\rangle\cdots|i\rangle|0\rangle,~~~ |0\rangle\cdots|i\rangle|i\rangle,\\
\vdots~~~~~~~~~~~~\vdots\\
|0\rangle|i\rangle\cdots|i\rangle|i\rangle,~|i\rangle\cdots|i\rangle|i\rangle.
\end{array}
\end{equation}

Hence, there is a total of $(d-1)$ subspace UPBs for different values of $i$. All these 
UPBs have one-dimensional overlap and, thus, are consistent with the total number of 
states $m(d-1)+1$ of the set $\mathbb{S}^\prime$. With respect to each UPB, 
residing in an $m$-qubit subspace, one can assign an $m$-partite bound entangled 
state and, following the earlier method, one can define a $(d-2)$-parameter family 
of bound entangled states. Furthermore, these subspace UPBs certify the local 
indistinguishability of the states within the set $\mathbb{S}^\prime$ and also the 
uncompletability in any locally extended Hilbert space.

\section{DISCRIMINATION PROTOCOL}\label{sec5}
In this section an entanglement-assisted local protocol is constructed to 
discriminate the states of the nonlocal completable set $\mathbb{S}$ as given 
in the previous section. Note that the normalization constants do not play any 
key role in the discrimination protocol. So, these constants are ignored for 
simplicity in this entire section. As mentioned earlier, the mathematical 
structure of LOCC is still not clear and hence it is difficult to prove local 
distinguishability of a given set unless one constructs an explicit local protocol. 
This is why construction of a local protocol is so important.

\begin{theorem}\label{th3}
A two-qubit, bipartite maximally entangled Bell state as a resource is sufficient to 
distinguish the states of the set $\mathbb{S}$ by means of local operations and 
classical communication. 
\end{theorem}

\begin{proof}
To prove the above, it is required to build a local protocol by which the discrimination 
of the states of the aforementioned set $\mathbb{S}$ is possible using a maximally entangled 
Bell state $|\phi^+\rangle$ = $(|00\rangle+|11\rangle)\in\mathbb{C}^2\otimes
\mathbb{C}^2$ as a resource. Assume that the resource state is shared between the 
$(m-1)$th and the $m$th party. So, the last two parties hold two qubits each. The 
states of the set $\mathbb{S}$ along with the resource state $|\phi^+\rangle$ are 
presented by

\begin{equation}\label{eq16}
\begin{array}{c}
|0\rangle\cdots|0\rangle(|i0\rangle|00 + i0\rangle + |i1\rangle|01 + i1\rangle), \\[1ex]

|0\rangle\cdots|0\rangle(|i0\rangle|00 - i0\rangle + |i1\rangle|01 - i1\rangle), \\[1ex]

|0\rangle\cdots|i\rangle(|00 +i0\rangle|00\rangle + |01 + i1\rangle|01\rangle), \\[1ex]

|0\rangle\cdots|i\rangle(|00 - i0\rangle|00\rangle + |01 - i1\rangle|01\rangle), \\[1ex] 

\vdots\\[1ex] 

|i\rangle|0 + i\rangle|0\rangle\cdots|0\rangle(|00\rangle|00\rangle + |01\rangle|01\rangle), 
\\[1ex] 

|i\rangle|0 - i\rangle|0\rangle\cdots|0\rangle(|00\rangle|00\rangle + |01\rangle|01\rangle), 
\\[1ex] 

|0 + i\rangle|0\rangle\cdots|0\rangle(|00\rangle|i0\rangle+|01\rangle|i1\rangle), \\[1ex]

|0 - i\rangle|0\rangle\cdots|0\rangle(|00\rangle|i0\rangle+|01\rangle|i1\rangle). \\[1ex]
\end{array}
\end{equation}

Next, the discrimination protocol is described step by step: (i) First of all, the $m$th 
party does a two-outcome projective measurement while corresponding measurement 
operators are given by - $M_1^{(m)}$ = $|00\rangle\langle00| + \sum_i|i1\rangle\langle 
i1|$ and $M_2^{(m)}$ = $|01\rangle\langle01| + \sum_i|i0\rangle\langle i0|$. After 
performing the measurement, if the measurement outcome is ``1,'' then the above set 
is transformed to the following set:

\begin{equation}\label{eq17}
\begin{array}{c}
|0\rangle\cdots|0\rangle(|i0\rangle|00\rangle + |i1\rangle|i1\rangle), \\[1ex] 

|0\rangle\cdots|0\rangle(|i0\rangle|00\rangle - |i1\rangle|i1\rangle), \\[1ex] 

|0\rangle\cdots|0\rangle|i\rangle|00 + i0\rangle|00\rangle, \\[1ex]

|0\rangle\cdots|0\rangle|i\rangle|00 - i0\rangle|00\rangle, \\[1ex]

\vdots\\[1ex]

|i\rangle|0 + i\rangle|0\rangle\cdots|0\rangle|00\rangle|00\rangle, \\[1ex] 

|i\rangle|0 - i\rangle|0\rangle\cdots|0\rangle|00\rangle|00\rangle, \\[1ex] 

|0 + i\rangle|0\rangle|0\rangle\cdots|0\rangle|01\rangle|i1\rangle, \\[1ex]

|0 - i\rangle|0\rangle|0\rangle\cdots|0\rangle|01\rangle|i1\rangle. \\[1ex]
\end{array}
\end{equation}

(ii) Then, the $(m-1)$th party performs a two-outcome projective measurement and 
the corresponding measurement operators are given by - $M_1^{(m-1)}$ = $|01\rangle
\langle01|$, $M_2^{(m-1)}$ = $|00\rangle\langle00| + \sum_i(|i0\rangle\langle i0| + 
|i1\rangle\langle i1|)$. After this measurement if the measurement outcome is ``1'' 
then the states of the last two rows of the previous equation get eliminated. These states 
can be distinguished further by following two easy steps: the $m$th party performs a 
$(d-1)$-outcome projective measurement on his (or her) system. For each measurement 
outcome, there are two orthogonal states remaining which can be distinguished perfectly 
by LOCC for sure according to Walgate {\it et al.} \cite{Walgate00}. 

(iii) Now, go back to step (ii) again. If the measurement outcome is ``2'' after the 
measurement by the $(m-1)$th party then the states of the first $2(m-1)$ rows are left. 
To distinguish these states, the first party does a projective measurement defined by two 
measurement operators, $M_1^{(1)}$ = $|0\rangle\langle0|$ and $M_2^{(1)}$ = 
$\sum_i|i\rangle\langle i|$. Due to this measurement if the measurement outcome is 
``1,'' then the states of $2(m-2)$ rows are left. If the outcome is ``2'' here then the 
first party again performs a $(d-1)$-outcome projective measurement and, for every 
outcome, two orthogonal states are to be distinguished further. After the first party, these 
measurements are also performed by next $(m-3)$ parties, that is, excluding the 
$(m-1)$th and $m$th parties. This completes the distinguishability of all the states 
except the states of the first two rows.

(iv) So, for the states of the first two rows, the $(m-1)$th party does a $(d-1)$-outcome 
measurement where the corresponding measurement operators are $M_i^{\prime(m-1)}$ 
= $|i0\rangle\langle i0|+|i1\rangle\langle i1|$. For each $i$, there are two orthogonal 
states remaining which can be distinguished further. 

(v) Next, go back to step (i) again. If the measurement outcome is ``2'' due to the 
measurement by the $m$th party then there is another set of states like the set given in 
Eq.~(\ref{eq17}). This set can be distinguished in the same way as described above. 
In this way, the protocol completes. 
\end{proof}

Recall that the resource state in the above protocol is shared between the $(m-1)$th 
and the $m$th party. But because of the symmetry present within the states of the set 
$\mathbb{S}$, it is really not important which pair of parties holds the resource state. 
From the above protocol it is also proved that the amount of entanglement required to 
accomplish the task of distinguishing the OPSs of $\mathbb{S}$ by LOCC does not depend 
on the dimension of the subsystems. Again, it does not depend on the number of parties 
either. So, in the above protocol, entanglement is employed more efficiently than a 
teleportation-based protocol. In this context, it is important to mention that a 
teleportation-based protocol to distinguish the OPSs of $\mathbb{S}$ requires 
$(m-1)\mbox{log}_{2}d$ 
ebits. However, Theorem \ref{th3} exhibits the first ever example where the OPSs of a 
completable set is distinguished so efficiently via an entanglement-assisted local protocol. 
From the dimensional point of view, the present resource is an optimal resource. But it 
is not known 
whether a two-qubit nonmaximally entangled state can be employed for the perfect 
discrimination of the states of $\mathbb{S}$ by LOCC or not. This particular fact has also 
been pointed out before in the context of distinguishing the states that belong to a UPB 
\cite{Cohen08}. Theorem \ref{th3} also depicts that as long as the task of distinguishing 
the states of the set $\mathbb{S}$ is concerned, there exists at least one separable operation 
which can accomplish this task perfectly and can be implemented by LOCC with the help 
of a two-qubit maximally entangled Bell state as resource.

From the above it is clear that for a given nonlocal COPB if the nonlocality of such a COPB 
is solely because of the states of the set $\mathbb{S}$ then that COPB can be perfectly 
distinguished by LOCC with the help of a two-qubit maximally entangled Bell state shared 
between any two spatially separated parties. Note that the present discrimination protocol 
does not work for all uncompletable sets of this paper. Thus, it is yet to be known which 
kind of entangled states are sufficient to distinguish the states of the uncompletable sets.

\section{CONCLUSION AND OPEN PROBLEMS}\label{sec6}
In this paper, different classes of nonlocal sets of pure orthogonal product states have 
been constructed for arbitrarily high-dimensional multipartite quantum systems. 
These constructions are important for a better understanding about the phenomenon 
{\it quantum nonlocality without entanglement}. In particular, the completable 
nonlocal sets give insight regarding the separable operations that are not 
implementable by LOCC. As useful by-products of present nonlocal sets, a class 
of COPBs has been introduced from which no state can be eliminated by performing 
orthogonality-preserving measurements and also a class of multipartite bound entangled 
states has been introduced. After the present constructions, one immediate open problem 
is to generalize these sets for any high-dimensional multipartite quantum systems where 
the parties do not hold the same dimensional subsystems. A local protocol has also been 
constructed to distinguish the product states of the completable sets using a two-qubit 
maximally entangled Bell state as a resource. Nevertheless, it is quite difficult to find out an 
optimal resource to distinguish the states of a given set of product states by LOCC. Here 
are two interesting open problems: The first is to find (if it is possible to construct) a 
nonlocal orthogonal 
product basis for which a teleportation-based protocol is an optimal protocol. Second, the 
amount of entanglement required to realize a nonlocal separable operation by LOCC is 
considered here while the task is fixed; that is, the state discrimination task. So, if it is possible 
to define a universal entanglement cost (task independent) of separable operation then it 
will be interesting. In a given Hilbert space, the sets constructed here are small sets, that is, 
the number of states contained in a set is much less than the net dimension of the Hilbert 
space. Hence, an essential search in this direction is to find out the number of orthogonal 
product states that is necessary to certify the fact that no state can be eliminated 
from a set in a given Hilbert space by performing orthogonality-preserving measurements. 
Explicit constructions of such sets are also desired for any multipartite quantum system.

\section*{ACKNOWLEDGMENT}
Part of this work was completed when the author was supported by fellowships 
from the Council of Scientific and Industrial Research, Government of India, 
and Bose Institute.

\end{document}